\documentstyle{article}
\newcommand{\Bbb}[1]{{\bf #1}}
\newcommand{\frak}[1]{{\bf #1}}
\newcommand{\Z}{{\Bbb Z}}

\newcommand{\C}{{\Bbb C}}

\newcommand{\Ref}[1]{{\rm{(}\ref{#1}\rm{)}}}

\newcommand{\ben}{\begin{equation}}
\newcommand{\een}{\end{equation}}
\newcommand{\bean}{\begin{eqnarray}}
\newcommand{\eean}{\end{eqnarray}}
\newcommand{\be}{\begin{displaymath}}
\newcommand{\ee}{\end{displaymath}}
\newcommand{\bea}{\begin{eqnarray*}}
\newcommand{\eea}{\end{eqnarray*}}
\newcommand{\g}{{{\frak g}\,}}

\newcommand{\h}{{{\frak h\,}}}

\newcommand{\Id}{{\rm Id}}

\newcommand{\Eta}{{\rm{H}}}

\newenvironment{definition}{\par\vspace{.5\baselineskip}
\noindent{\it Definition\/}:}{\par\vspace{.5\baselineskip}}
\newtheorem%
{thm}{Theorem}[section]
\newtheorem%
{proposition}[thm]{Proposition}
\newtheorem%
{lemma}[thm]{Lemma}
\newtheorem%
{corollary}[thm]{Corollary}
\newcommand{\End}{{\rm{End}}}
\newcommand{\RR}[3]{R^{(#1#2)}(z_{#1#2},\lambda#3)}
\author{Giovanni Felder\thanks{Supported in part by NSF grant
DMS-9400841}
\\
Forschungsinstitut f\"ur Mathematik,
ETH-Zentrum,\\
CH-8092 Z\"urich, Switzerland\\
and\\
Department of Mathematics,\\
University of North Carolina at Chapel Hill,\\
Chapel Hill, NC 27599-3250, USA%
\thanks{Permanent address}
}
\title{Conformal field theory and integrable systems\\
associated to
elliptic curves}
\date{July 1994}
\begin{document}
\maketitle
\section{Introduction}

It has become clear over the years that quantum groups
(i.e., quasitriangular Hopf algebras, see \cite{D})
and their semiclassical counterpart,
Poisson Lie groups, are an essential algebraic
structure underlying three related subjects:
integrable models of statistical mechanics,
conformal field theory and integrable models
of quantum field theory in 1+1 dimensions.
Still, some points remain obscure form the point
of view of Hopf algebras. In particular,
integrable models associated with elliptic
curves are still poorly understood. We propose
here an elliptic version of quantum groups, based on
the relation to conformal field theory, which
hopefully will be helpful to complete the picture.

But before going to the elliptic case, let us
remind the relations between the three subjects
in the simpler rational and trigonometric
cases.

In integrable models  of statistical mechanics (see
\cite{Ba},\cite{F}), the basic
object is an $R$-matrix, i.e., a meromorphic
function of a spectral parameter
$z\in\C$ with values in $\End(V\otimes V)$ for some
vector space $V$, obeying of the
Yang--Baxter equation
\be\label{YBE}
R^{(12)}(z)
R^{(13)}(z+w)
R^{(23)}(w)
=
R^{(23)}(w)
R^{(13)}(z+w)
R^{(12)}(z),
\ee
in $\End(V\otimes V\otimes V)$. The notation is customary in
this subject: $X^{(j)}\in\End(V\otimes\cdots\otimes V)$,
 for $X\in\End(V)$, means
$\Id\otimes
\cdots\otimes
\Id\otimes X\otimes \Id\otimes\cdots\otimes\Id$, with $X$ at
the $j$th place, and if $R=\Sigma X_\nu\otimes Y_\nu$,
$R^{(ij)}=\Sigma X_\nu^{(i)}Y_\nu^{(j)}$.

The Yang--Baxter equation implies the commutativity of infinitely many
transfer matrices constructed out of $R$. Rational
and trigonometric solutions of the Yang--Baxter equation
appear naturally in the theory of Quasitriangular Hopf
algebras.

If $R$ depends on a parameter $\hbar$ so that $R=\Id +\hbar r
+O(\hbar^2)$, as $\hbar\to 0$, then the ``classical
$r$-matrix'' $r$ obeys the classical Yang--Baxter equation
\ben\label{CYBE}
[r^{(12)}(z),r^{(13)}(z+w)+r^{(23)}(w)]+[r^{(13)}(z+w),
r^{(23)}(w)]=0.
\een
This equation appears in the theory of Poisson--Lie groups,
but has the following relation with conformal field
theory. In the skew-symmetric case $r(z)=-r^{(21)}(-z)$,
it is the compatibility condition for the system of equations
\ben\label{KZ}
\partial_{z_i}u=\sum_{j:j\neq i}r^{(ij)}(z_i-z_j)u
\een
for a function $u(z_1,\dots,z_n)$ on $\C^n-\cup_{i<j}\{z|z_i=z_j\}$,
with values in $V\otimes\cdots\otimes V$. In the rational case,
very simple skew-symmetric solutions are known:
$r(z)=C/z$, where $C\in\g\otimes\g$ is a symmetric
invariant tensor of a finite dimensional Lie algebra
$\g$ acting on a representation space $V$. The corresponding
system of differential equations is the Knizhnik--Zamolodchikov (KZ)
equation for conformal blocks of the Wess--Zumino--Witten
model of conformal field theory on the sphere. Solutions
of \Ref{CYBE} in $\g\otimes\g$, for simple Lie
algebras $\g$, were partially classified by Belavin and Drinfeld
\cite{BD}, who in particular proved that, under a non-degeneracy
assumption, solutions can be divided into three classes,
according, say, to the lattice of poles: rational, trigonometric
and elliptic. Elliptic solutions are completely classified
and exist only for $sl_N$.

Recently, Frenkel and Reshetikhin \cite{FR} considered
the ``quantization'' of the Knizhnik--Zamolodchikov equations
based on the representation theory of Yangians (rational case)
and affine quantum enveloping algebras (trigonometric case).
They are a compatible system of {\em difference} equations which
as $\hbar\to 0$ reduce to the differential equations \Ref{KZ}.
In an important special case, these difference equations
had been introduced earlier by Smirnov
\cite{S} who derived them as equations for ``form factors''
in integrable quantum field theory, and gave relevant solutions.
In the quantum field
theory setting $R$ has the interpretation of two-particle
scattering matrix, and is required to obey the ``unitarity''
relation $R(z)R^{(21)}(-z)=\Id$, as well as a
``crossing symmetry'' condition.

In the elliptic case, one knows solutions of the Yang--Baxter
equation whose semiclassical limit are the $sl_N$ solutions
discussed above \cite{Bel}. The relevant algebraic structure
is here the Sklyanin algebra \cite{S}, \cite{Ch},
which however does not fall into
the general quantum group theory.
Although  elliptic solutions related to other Lie algebras
have not been found (and they could not have a semiclassical
limit by the Belavin--Drinfeld theorem), many elliptic solutions
of the Star-Triangle relation, a close cousin of the
Yang--Baxter equation, are known (see \cite{JMO},
\cite{JKMO}, \cite{DJKMO}). This is somewhat
mysterious, as, in the trigonometric case, solutions of
both equations are in one-to-one correspondence.
Another apparent puzzle we want to point out, is that conformal
field theory can be defined on arbitrary Riemann surfaces
\cite{TUY}  whereas $r$-matrices exist only up to genus one,
and in genus one only for $sl_N$.

We will start from the solution to this last puzzle to
arrive, via quantization and difference equations to
a notion of elliptic quantum group, which is the
algebraic structure underlying the elliptic
solutions of the Star-Triangle relation.

Let us mention some other recent progress in
similar directons. In \cite{FR} solutions of
the Star-Triangle relation are obtained as
connection matrices for the trigonometric
quantum KZ equations. Very recently,
Foda et al.\ \cite{FIJKMY} have proposed an elliptic quantum
algebra of non-zero level, by another
modification of the Yang--Baxter
equation.

\vskip 10pt
\noindent{\it Acknowledgment.} I wish to thank V. Kac,
who asked a question that triggered the quantum
part of this research.

\section{Conformal field theory, KZB equations}
Our starting point is the set of genus one
Kni\-zhnik--Za\-mo\-lo\-dchi\-kov--Ber\-nard
(KZB) equations,
obtained by Bernard \cite{Be1,Be2}
as generalization of the KZ equations. These equations have
been studied recently in
\cite{FaGa}, \cite{EK}, \cite{FW}.

Let $\g$ be a simple complex
Lie algebra with invariant bilinear
form normalized in such a way that long roots
have square length 2. Fix a Cartan subalgebra $\h$.
 The KZB equations are equations
for a  function $u(z_1,\dots,z_n,\tau,\lambda)$ with values
in the weight zero subspace
(the subspace killed by $\h$)
of a tensor product of irreducible finite dimensional
 representations of $\g$. The arguments
$z_1,\dots z_n,\tau$ are comlex numbers with $\tau$ in the
upper half plane, and the $z_i$ are distinct modulo the
lattice $\Z +\tau\Z$, and $\lambda\in\h$. Let us introduce
coordinates $\lambda=\Sigma\lambda_\nu h_\nu$ in terms
of an orthonormal basis ($h_\nu$) of $\h$.
In the formulation of \cite{FW}, the KZB equations
 take the form
\begin{eqnarray}\label{KZB}
\kappa\partial_{z_j}u&=&
-\sum_\nu h_\nu^{(j)}\partial_{\lambda_\nu}u
+\sum_{l:l\neq j}
\Omega^{(j,l)}(z_j-z_l,\tau,\lambda)u,\\
4\pi i\kappa\partial_\tau u&=&
\sum_\nu\partial_{\lambda_\nu}^2u
+\sum_{j,l}
\Eta^{(j,l)}(z_j-z_l,\tau,\lambda)u,\label{KZBTAU}
\end{eqnarray}
Here $\kappa$ is an integer parameter which is large
enough depending on the representations in the tensor
product and $\Omega$, $\Eta\in\g\otimes\g$ are tensors
preserving the weight zero subspace that we now
describe. Let $\g=\h+\sum_{\alpha\in\Delta}\g_\alpha$ be
the root decompostion of $\g$, and $C\in S^2\g$ be the
symmetric invariant tensor dual to the invariant bilinear
form on $\g$. Write $C=\sum_{\alpha\in\Delta\cup\{0\}}C_\alpha$,
where $C_0=\sum_\nu h_\nu\otimes h_\nu$ and $C_\alpha\in
\g_\alpha\otimes\g_{-\alpha}$.
 Let $\theta_1(t,\tau)$ be Jacobi's
theta function
\be
\theta_1(t|\tau)=-\sum_{j=-\infty}^{\infty}
e^{\pi i(j+\frac12)^2\tau+2\pi i(j+\frac12)(t+\frac12)}.
\ee
and introduce functions $\rho$, $\sigma$:
\begin{eqnarray*}
\rho(t)&=&\partial_t\log\theta_1(t|\tau),\\
\sigma(w,t)&=&
\frac{\theta_1(w-t|\tau)\partial_t\theta_1(0|\tau)}
{\theta_1(w|\tau)\theta_1(t|\tau)}.
\end{eqnarray*}
The tensor $\Omega$ is given by
\be
\Omega(z,\tau,\lambda)=
\rho(z)C_0+\sum_{\alpha\in\Delta}\sigma(\alpha(\lambda),z)C_\alpha
\ee
The tensor $\Eta$ has a similar form.
We need the following special functions of $t\in\C$, expressed
in terms of $\sigma(w,t)$, $\rho(t)$ and Weierstrass'
elliptic function $\wp$  with periods $1,\tau$.
\begin{eqnarray*}
I(t)&=&\frac12(\rho(t)^2-\wp(t)),\\
J_w(t)&=&\partial_t\sigma(w,t)+(\rho(t)+\rho(w))\sigma(w,t).
\end{eqnarray*}
These functions are regular at $t=0$. The tensor $\Eta$ is
then given by the formula
\be
\Eta(t,\tau,\lambda)=I(t)C_0+\sum_{\alpha\in\Delta}
J_{\alpha(\lambda)}(t)C_\alpha.
\ee
As shown in \cite{FW}, the functions $u$ from conformal field
theory have a special dependence on the parameter $\lambda$.
For fixed $z,\tau$, the function $u$, as a function of $\lambda$,
 belongs to a finite dimensional
space of antiinvariant
theta function of level $\kappa$. Therefore the right way
of looking at these equation is to consider $u$ as a function
of $z_1,\dots, z_n,\tau$ taking values in a finite dimensional
space of functions of $\lambda$.

The tensors have the skew-symmetry property
 $\Omega(z)+\Omega^{(21)}(-z)=0$,
and $\Eta(z)-\Eta^{(21)}(-z)=0$ and commute with
$X^{(1)}+X^{({2})}$ for all $X\in\h$.
The compatibility condition of \Ref{KZB} is then the
{\em modified classical Yang--Baxter  equation}
 \cite{FW}
\begin{eqnarray}\label{MCYBE}
\sum_\nu
\partial_{\lambda_\nu}\Omega^{(1,2)}h_\nu^{(3)}+
\sum_\nu\partial_{\lambda_\nu}\Omega^{(2,3)}h_\nu^{(1)}+
\sum_\nu\partial_{\lambda_\nu}\Omega^{(3,1)}h_\nu^{(2)} & &
\nonumber
\\
-[\Omega^{(1,2)},\Omega^{(1,3)}]
-[\Omega^{(1,2)},\Omega^{(2,3)}]
-[\Omega^{(1,3)},\Omega^{(2,3)}] &=& 0
\end{eqnarray}
in $\g\otimes\g\otimes\g$.
In this equation, $\Omega^{(ij)}$ is taken at $(z_i-z_j,\tau,\lambda)$.
Moreover, there are relations involving $\Eta$, which we do
not consider here, as we will consider only the first equation
\Ref{KZB}.
The quantization of \Ref{KZBTAU} is an important open problem
related, for $n=1$, to the theory of elliptic Macdonald polynomials
\cite{EK}.

\section{The quantization}
 Let $\h$ be the complexification of a Euclidean space $\h_r$ and
extend the scalar product to a bilinear form on $\h$. View $\h$ a an
Abelian Lie algebra. We consider finite dimensional diagonalizable
$\h$-modules $V$. This means that we have a weight decomposition
$V=\oplus_{\mu\in\h}V[\mu]$ such that $\lambda\in\h$ acts as
$(\mu,\lambda)$ on $V[\mu]$.  Let $P_\mu\in\End(V)$ be the projection
onto $V[\mu]$.

It is convenient to introduce the following notation.
Suppose $V_1,\dots, V_n$ are finite dimensional diagonalizable
$\h$-modules. If $f(\lambda)$ is a meromorphic function on $\h$ with
values in $\otimes_iV_i=V_1\otimes\cdots\otimes V_n$
or $\End{(\otimes_iV_i)}$, and $\eta_i$ are
complex numbers, we define a function on $\h$
\be
 f(\lambda+\sum\eta_ih^{(i)})=\sum_{\mu_1,\dots,\mu_n}
\prod_{i=1}^nP_{\mu_i}^{(i)}f(\lambda+\Sigma\eta_i\mu_i),
\ee
taking values in the same space as $f$.

Given $\h$ and $V$ as above,
the quantization of \Ref{MCYBE} is  an equation
for a meromorphic function $R$ of the spectral parameter
$z\in\C$ and an additional variable $\lambda\in\h$,
taking values in $\End(V\otimes V)$
\begin{eqnarray}
 &\RR 12{+\eta h^{(3)}}
\RR 13{-\eta h^{(2)}}
\RR 23{+\eta h^{(1)}}=& \nonumber\\
 &=
\RR 23{-\eta h^{(1)}}
\RR 13{+\eta h^{(2)}}
\RR 12{-\eta h^{(3)}}.&
\label{MYBE}\end{eqnarray}
The parameter $\eta$ is proportional to $\hbar$,
and $z_{ij}$ stands for $z_i-z_j$.
This equation forms the basis for the subsequent
analysis. Let us call it modified Yang--Baxter equation
(MYBE). Note that a similar equation,
without spectral parameter, has appeared for the monodromy
matrices in Liouville theory, see
\cite{GN}, \cite{Ba}, \cite{AF}.
We supplement it by the ``unitarity'' condition
\ben
\label{UC}
\RR 12{}\RR 21{}={\rm{Id}}_{V\otimes V},
\een
 and the ``weight zero''
condition
\ben\label{WZ}
[X^{(1)}+X^{(2)},R(z,\lambda)]=0,\qquad \forall X\in\h.
\een
We say that  $R\in\End(V\otimes V)$ is a {\em generalized quantum $R$-matrix}
if it obeys \Ref{MYBE}, \Ref{UC}, \Ref{WZ}.

If we have a family of solutions parametrized by $\eta$ in
some neighborhood of the origin, and $R(z,\lambda)=
{\rm Id}_{V\otimes V}-2\eta \Omega(z,\lambda) +O(\eta^2)$ has
a ``semiclassical asymptotic expansion'', then \Ref{MYBE}
reduces to the modified classical Yang--Baxter
equation \Ref{MCYBE}.

Here are examples of solutions. Take $\h$ to be the
Abelian Lie algebra of
diagonal $N$ by $N$ complex matrices, with bilinear
form Trace($AB$),  acting on $V=\C^N$. Denote by $E_{ij}$ the
$N$ by $N$ matrix with a one in the $i$th row and $j$th column
and zeroes everywhere else. Then we have

\begin{proposition}
The function
\be\label{sol}
R(z,\lambda)=\sum_{i}E_{ii}\otimes E_{ii}
+\sum_{i\neq j}
\frac{\sigma(\gamma,\lambda_{ij})}
{\sigma(\gamma,z)}
E_{ii}\otimes E_{jj}
+\sum_{i\neq j}
\frac{\sigma(\lambda_{ij},z)}
{\sigma(\gamma,z)}
E_{ij}\otimes E_{ji},
\ee
is a ``unitary'' weight zero
solution of the modified Yang--Baxter equation, i.e., it
is a generalized quantum $R$-matrix,
with
$\eta=\gamma/2$.
\end{proposition}

\noindent The proof is based on comparing poles and behavior
under translation of spectral parameters by $\Z+\tau\Z$ on
both sides of the equation.
It uses unitarity, the $\Z$ periodicity of $R$ and the
transformation property
\bean\label{TAU}
R(z+\tau,\lambda)&=&
e^{-4\pi i\eta}\exp(2\pi i ( \eta C_0+\lambda^{(1)}))
R(z,\lambda)\exp(2\pi i(\eta C_0-\lambda^{(1)})),\\
C_0&=&\sum E_{ii}\otimes E_{ii}.
\eean
Two limiting cases of this solution are of interest.
First, if $\tau\to i\infty$ and $\lambda_j-\lambda_l\to i\infty$,
if $j<l$, we recover the well known trigonometric
$R$ matrix connected with the quantum enveloping algebra
of $A_{N-1}^{(1)}$ (see \cite{J}).

The semiclassical limit is more subtle. To obtain precisely $\Omega$
of the KZB equations, replace $\sigma(\gamma,\lambda_{ij})$ by
$\sigma(\gamma,\lambda_{ij})\exp(\rho(\lambda_{ij})\gamma)$.  It turns
out that this replacement is compatible with the MYBE
(but violates the assumption of meromorphy). Then $-\gamma R$
has a semiclassical asymptotic expansion with $\Omega$ (for $gl_N$)
as coefficient
of $-\gamma$.

Following the Leningrad school (see \cite{F}),
one associates a
bi\-al\-ge\-bra with qua\-dra\-tic relations
to each solution of the Yang--Baxter equation.
In our case we have modify slightly the construction. Let us
consider an ``algebra'' $A(R)$ associated to a generalized
quantum $R$-matrix $R$, generated by meromorphic functions
on $\h$ and the matrix elements (in some
basis of $V$) of a matrix $L(u,\lambda)\in\End(V)$ with
non commutative entries, subject to the relations
\bea &
R^{(12)}(z_{12},\lambda+\eta h)
L^{(1)}(z_1,\lambda-\eta h^{(2)})
L^{(2)}(z_2,\lambda+\eta h^{(1)})=& \\
&=
L^{(2)}(z_2,\lambda-\eta h^{(1)})
L^{(1)}(z_1,\lambda+\eta h^{(2)})
R^{(12)}(z_{12},\lambda-\eta h).&
\eea
Instead of giving a more precise definition of this algebra, let us
define the more important notion (for our purposes) of representation
of $A(R)$.

\begin{definition} Let $R\in\End(V\otimes V)$
be a meromorphic unitary weight zero solution of the MYBE
(a generalized quantum $R$-matrix). A
representation of $A(R)$ is a diagonalizable $\h$-module $W$ together
with a meromorphic function $L(u,\lambda)$ (called $L$-operator)
on $\C\times\h$ with values
in $\End(V\otimes W)$ such that the identity
\bea
&R^{(12)}(z_{12},\lambda+\eta h^{(3)})
L^{(13)}(z_1,\lambda-\eta h^{(2)})
L^{(23)}(z_2,\lambda+\eta h^{(1)})=&
\\
&=
L^{(23)}(z_2,\lambda-\eta h^{(1)})
L^{(13)}(z_1,\lambda+\eta h^{(2)})
R^{(12)}(z_{12},\lambda-\eta h^{(3)})&
\eea
holds in $\End(V\otimes V\otimes W)$, and so that
$L$ is of weight zero:
\be
[X^{(1)}+X^{(2)},L(u,\lambda)]=0, \qquad \forall X\in\h.
\ee
We have natural notions of homomorphisms of representations.
\end{definition}

\begin{thm}\label{t1} (Existence and coassociativity of
the coproduct) Let $(W,L)$ and $(W',L')$  be representations
of $A(R)$. Then $W\otimes W'$ with $\h$-module
structure $X(w\otimes w')=Xw\otimes w'+w\otimes Xw'$ and
$L$-operator
 \be
L^{(12)}(z,\lambda+\eta h^{(3)})
L^{(13)}(z,\lambda-\eta h^{(2)})
\ee
is a representation of $A(R)$.
 Moreover, if we have three
representation $W$, $W'$, $W''$, then the representations
$(W\otimes W')\otimes W''$ and $W\otimes(W'\otimes W'')$ are
isomorphic (with the obvious isomorphism).
\end{thm}

\noindent Note also that if $L(z,\lambda)$ is an $L$-operator then
also $L(z-w,\lambda)$ for any complex number $w$.
Since the MYBE and the weight zero condition
mean that $(V,R)$ is a representation,
we may construct representations on $V^{\otimes n}=
V\otimes\cdots\otimes
V$ by iterating the construction of Theorem \ref{t1}.
The corresponding $L$ operator is the ``monodromy matrix''
with parameters $z_1,\dots,z_n$:
\be
\prod_{j=2}^{n+1}R^{(1j)}(z-z_j,\lambda-\eta
\Sigma_{1<i<j}h^{(i)}+\eta\Sigma_{j<i\leq n+1}h^{(i)}).
\ee
(the factors are ordered from left to right). Although
the construction is very reminiscent of the Quantum Inverse
Scattering Method \cite{F}, we cannot at this point
construct commuting transfer matrices by taking the trace of the
monodromy matrices. As will be explained below, one
has to pass to IRF models.

\section{Difference equations}
We now give the quantum version of the KZB equation \Ref{KZB}. As
in the trigonometric case \cite{S, FR} it is a system of
difference equations. The system is symmetric, i.e., there
is an action of the symmetric group mapping solutions to
solutions. In the trigonometric and rational case, symmetric
meromorphic solutions with proper  pole structure are
``form factors'' of integrable models of Quantum Field Theory
in two dimensions \cite{S}.

It is convenient to formulate the construction in terms of
representation theory of the affine symmetric group. Let
$S_n$ be the symmetric group acting on $\C^n$ by permutations of
coordinates, and $s_j$, $j=1,\dots,n-1$ be the transpositions
$(j,j+1)$.  These transpositions generate $S_n$ with relations
$s_js_l=s_ls_j$, if $|j-l|\geq 2$, $s_js_{j+1}s_j=s_{j+1}s_js_{j+1}$,
and $s_j^2=1$. Let also $P\in\End(V\otimes V)$ be the ``flip'' operator
$P u\otimes v=v\otimes u$ and if $R$ is a generalized quantum $R$-matrix,
set $\hat R=RP$. The defining properties of a generalized quantum $R$-matrix
imply:

\begin{proposition} Suppose that $R$ is a generalized quantum $R$-matrix.
The formula
\ben\label{SG}
s_jf(z,\lambda)=\hat \RR j{,j+1}{-\eta
\Sigma_{i<j}h^{(i)}+\eta\Sigma_{i>j+1}h^{(i)}}
f(s_jz,\lambda)
\een
defines a representation of $S_n$ on meromorphic
functions on $\C\times\h$ with values in $V^{\otimes n}$.
\end{proposition}

\noindent The (extended) affine symmetric group $S_n^a$ is
the semidirect product of $S_n$ by $Z_n$. It is generated
by $s_j$ and commuting generators $e_j$, $j=1,\dots,n$,
with relations $s_je_l=e_ls_j$, if $l\neq j,j+1$ and
$s_je_j=e_{j+1}s_j$. Let us introduce a parameter $a\in\C$
and let $e_j$ act on $z\in\C^n$ as $e_j(z_1,\dots,z_n)=
(z_1,\dots,z_j-a,\dots,z_n)$. Note that $S_n^a$ is actually
generated by $s_1$, \dots, $s_{n-1}$ and $e_n$, since
the other $e_j$ are constructed recursively as
$e_j=s_je_{j+1}s_j$.

\begin{thm} Suppose that $R$ is a generalized quantum $R$-matrix.
Let $T_jf(z,\lambda)$ $=f(z,\lambda-2\eta h^{(j)})$,
$\Gamma_jf(z,\lambda)=f(z+a,\lambda)$ and
$R^{(j,n)}$ denote the operator of multiplication
by $\RR j{,n}{-\eta\Sigma_{i<j}h^{(i)}+\eta\Sigma_{j<i<n}h^{(i)}}$.
Then \Ref{SG} and
\be
e_nf=R^{(n-1,n)}\cdots R^{(2,n)}R^{(1,n)}\Gamma_nT_nf
\ee
define a representation of $S_n^a$ on meromorphic
functions on $\C\times\h$ with values in $V^{\otimes n}$.
\end{thm}

\noindent It is easy to calculate the action of the other generators
$e_j$. One gets expressions similar to the ones in
\cite {FR,S}.

The compatible system of difference equations
(Quantum KZB equations) is then
\be
e_jf=f,\qquad j=1,\dots n.
\ee
The symmetric group maps solutions to solutions.

Moreover, it turns out that, for special values of $a$,
the representation of $S_n^a$ for the solution of Prop.\ \ref{sol}
preserves a space of theta functions, as in the classical case.

\section{IRF models}
In our setting, the relation between the generalized quantum $R$-matrix and
the Boltzmann weights $W$ of the corresponding
interaction-round-a-face (IRF)
model \cite{Ba} is very simple. Let $R\in\End(V\otimes V)$
be a generalized quantum $R$-matrix, and let $V[\mu]$ be the component
of weight $\mu\in \h^*$ of $V$, with projection
$E[\mu]:V\to V[\mu]$. Then for $a,b,c,d\in\h^*$, such
that  $b-a$, $c-b$, $d-a$ and $c-d$ occur in the weight
decomposition of $V$, define
a linear map
\be
W(a,b,c,d,z,\lambda):V[d-a]\otimes V[c-d]\to
V[c-b]\otimes V[b-a],
\ee
by the formula
\ben
W(a,b,c,d,z,\lambda)=E[c-b]\otimes E[b-a]
R(z,\lambda-\eta a-\eta c)|_{V[d-a]\otimes V[c-d]}.
\een
Note that $W(a+x,b+x,c+x,d+x,z,\lambda+2\eta x)$ is independent
of $x\in\h\simeq\h^*$. Set $W(a,b,c,d,z)=W(a,b,c,d,z,0)$.

\begin{thm}
If $R$ is a solution of the MYBE,
 then $W(a,b,c,d,z)$ obeys the Star-Triangle relation
\bea
&\sum_g
W(b,c,d,g,z_{12})^{(12)}
W(a,b,g,f,z_{13})^{(13)}
W(f,g,d,e,z_{23})^{(23)}&
\\
&=
\sum_g
W(a,b,c,g,z_{23})^{(23)}
W(g,c,d,e,z_{13})^{(13)}
W(a,g,e,f,z_{12})^{(12)},&
\eea
on $V[f-a]\otimes V[e-f]\otimes V[d-e]$.
\end{thm}
The familiar form of the Star-Triangle relation \cite{Ba},\cite{JMO}
is recovered when the spaces $V[\mu]$ are 1-dimensional. Upon
choice of a basis, the
Boltzmann weights $W(a,b,c,d,z)$ are then numbers.

For example, if $R$ is the solution of Prop.\ \ref{sol}, we obtain
the well-known $A^{(1)}_{n-1}$ solution (see \cite{JMO},
\cite{JKMO} and references
therein).

 It is known that solutions of the Star-Triangle relations
can be used to construct solvable models of statistical mechanics.
Several elliptic solutions are known. It is to
be expected that the representation theory of the algebra
$A(R)$ above will give a more systematic theory of
solutions.
Also, the fact that these Boltzmann weights arise as
connection matrices of the quantum KZ equation \cite{FR}
and the similarity of our Yang--Baxter equation
with the triangle equation of \cite{GN}, suggest
that our algebra is the quantum conformal field
theory analogue of $U_q(\g)$, the algebra governing
the monodromy of conformal field theory.

\end{document}